\documentstyle[12pt,epsf]{l-aa}                                                    

\begin{document}
\setcounter{footnote}{1}

\thesaurus{011      % Section 01: Letters
     (08.14.1; % Stars: neutron                                             
      13.07.1)}% Gamma rays: bursts 
     
\title{An independent estimate of the cosmological distance to GRB970228
and GRB970508}

\author{V.M.Lipunov\inst{1,2}
K.A.~Postnov\inst{1,2}
\and M.E.~Prokhorov\inst{2}}

\institute{
Faculty of Physics, Moscow State University,
\and Sternberg Astronomical Institute, Moscow State University,
                119899 Moscow, Russia
}
\date{Received ... 1997, accepted ..., 1997}
\maketitle
\markboth{V.Lipunov,K.Postnov \& M.Prokhorov. GRB970228}{ ...}

\begin{abstract}

Assuming binary neutron star mergings as a standard-candle model for
GRBs, an independent estimate is obtained for the redshift of
GRB970228 and GRB970508 with optical counterparts, using mean
statistical properties of GRBs observed by BATSE. We derive $z=0.7\pm
0.1$ and $z=1.9\pm 0.1$ for GRB970228 and GRB970508 respectively,
depending on the power-law index of the GRB spectrum $s=-1.1\pm 0.3$
and the value of the redshift of the initial star formation $z_*=3-10$
in a flat $\Omega=1$ Universe with a cosmological term $\Lambda=0.7$.


\end{abstract}

\noindent
A fading optical transient object (Groot et al. 1997a) has been found
within gamma-ray and X-ray error boxes of GRB970228.  Deep optical
studies of this field with large telescopes revealed an elongated
($1^{\prime\prime}\times 1.5^{\prime\prime}$) faint ($R\simeq 24$)
object coincident with the position of an optical transient source
detected about one day after the gamma-ray burst (Groot et al. 1997b;
Metzger et al. 1997b).  X-ray spectra of the X-ray transient source
accompanied GRB970228, obtained by BeppoSAX (Costa et al. 1997b),
display soft X-ray absorption corresponding to the column density
toward the source $N_{HI}=1.6\times 10^{21}$ cm$^{-2}$ (which is in
agreement with a galactic absorption in this direction of $1.44\times
10^{21}$ cm$^{-2}$, as derived from neutron hydrogen emission
(V.G.Kurt, private communication).  Subsequent optical observations of
the optical transient source with the Hubble Space Telescope (HST) on
March 26, 1997, confirmed the presence of a faint star-like source
($I\approx 24.2$) embedded into a dim nebula (Sahu et al. 1997).  This
makes GRB970228 the first GRB identified with some astronomical
object, very likely a distant galaxy, and lends direct support to the
cosmological origin of GRBs.  Incidentally, one of the faint galaxies
within the X-ray error box at $1'$ distance from the optical transient
has a redshift $z=0.498$ (Metzger et al. 1997a). However, the redshift
of the faint nebula associated with the optical transient has not been
measured so far. More recent weaker GRB970508 is also associated with
a variable optical ($R\sim 20.5^m$) counterpart (Bond 1997), with a
redshift of $0.835<z<2.1$ (Metzger et al. 1997c), as measured by Fe
and Mg absorption lines, the upper limit being due to the absence of
Lyman-alpha forest in its spectrum.

Here we estimate of the redshift of GRB970228 and GRB970508 using the
mean statistical properties of observed GRBs.  We assume the
cosmological origin of GRBs as standard-candle binary neutron star
mergers.  The peak gamma-ray flux from GRB970228 and GRB970508
measured at BeppoSAX Gamma Ray Burst Monitor (GRBM) was 3700 counts/s
(Costa et al. 1997a) and 450 counts/s (Costa et al. 1997c), roughly
corresponding to a 60-600 keV flux of 6 and 0.7 photons cm$^{-2}$
s$^{-1}$ respectively, assuming the conversion factor quoted in Piro
et al. (1996) (note that this rate may be lower limits considering the
off-set positions of the outbursts).  Assuming rough correspondence of
the BeppoSAX and BATSE photon counts (which would not be too far-off),
we find the location of GRB970228 and GRB970508 on the $\log N-\log
F_{peak}$ curve for 3B BATSE GRB catalog (256 ms channel)
(Fig. 1). This curve can be fitted within the framework of
cosmological model of GRBs as coalescing binary neutron stars (see
Lipunov et al.  1995 for more detail) using: 1) cosmological model
parameters (the total density in units of the critical density to
close the Universe, $\Omega$, the density of baryons, $\Omega_b$, and
cosmological constant term, $\Omega_\Lambda$); 2) evolutionary
parameters (the fraction of elliptical galaxies among the total number
of galaxies, $\epsilon$, the redshift of the initial star formation,
$z_*$). Assuming $\Omega=1$, $\Omega_b=0.0046$, $\Omega_\Lambda=0.7$,
$\epsilon=0.5$, and approximating effective GRB spectrum as a single
power law with a photon index $s=-1.1\pm 0.3$ (Mallozzi et al. 1996),
we calculate the best-fit of the 3B $\log N-log F_{peak}$ distribution
for $s=-1.1$ and different values of $z_*=3,5,10$ (Fig. 1, solid
lines). The observed distribution is best-fitted by models with
$z_*>5$.  The vertical arrows indicate the positions of GRB970228 and
GRB970508.

This procedure allows us to determine the redshifts GRB970228 and
GRB970508 would have in models with different $z_*$ and $s$
(Table~1). The mean redshift of GRB970228 is $\langle z\rangle =
0.7\pm0.1$, and taht of GRB970508 is $z\sim 1.9\pm 0.1$ the error
being formally due to variations in the spectral slope and $z_*$.  The
dependence of the redshift on the countrate observed is shown in
Fig. 2 for various $z_*=3,5,10$. If this model is correct, the
brightest BATSE GRBs should be observed from redshifts $z_{min}\sim
0.1$.  Assuming this GRB to be not selected in its luminosity, the
redshift of the host galaxy is thus predicted to be about 0.7.

To conclude, we stress that the obtained redshift estimation of the
possible host galaxy of GRB970228 $z= 0.7$ and its variance $\pm 0.1$
assumes the hypothesis of the standard candle. From this point of
view, the direct measurement of the redshift of the nebula associated
with the optical transient would be a crucial test of the cosmological
origin of GRB970228 and is highly desirable. Our estimate of the
redshift of GRB970508 $z=1.9\pm 0.1$ falls within the limits obtained
from the spectroscopy of the optical counterpart $0.835<z<2.1$.

The authors acknowledge Profs. N.I.Shakura and V.G.Kurt
for valuable discussions. The work is partially supported by
Russian Fund for Basic Research through Grant No 95-02-06053-a.

\begin{table}
\caption{Redshift of GRB970228 for different
initial star formation redshifts $z_*$ and
gamma-ray spectral photon index $s$}
\begin{center}
\begin{tabular}{rrrr}
\hline
$z_*$ & \multicolumn{3}{c}{$z$} \\ \cline{2-4}
      & $s=1.4$ & $1.1\,$  & $0.8\,$ \\
\hline
 10   &   0.79  & 0.80 & 0.80 \\
  5   &   0.70  & 0.80 & 0.73 \\
  3   &   0.64  & 0.64 & 0.57 \\
\hline
\end{tabular}
\end{center}
\end{table}

\begin{figure}
\epsfxsize=\hsize
\epsfbox{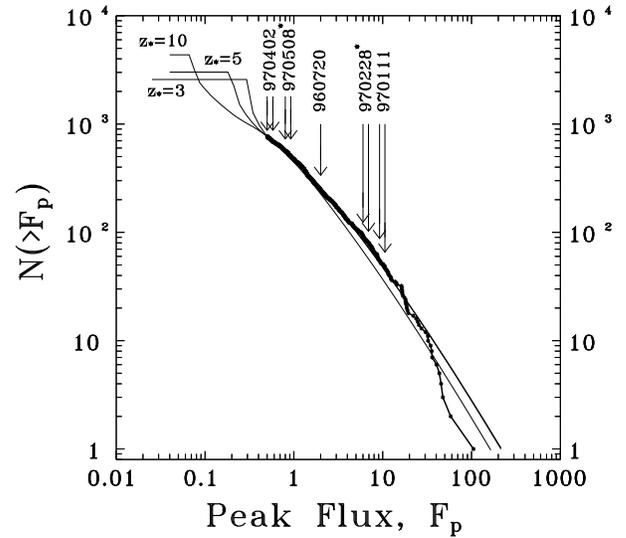}
\caption{
The $\log N-\log F_{peak}$ distribution of 3B BATSE GRBs from
256-ms 1-3 (50-300 keV) channels fitted with the cosmological model
distributions in a flat $\Omega=1$ Universe with a cosmological term
$\Omega_\Lambda=0.7$ assuming gamma-ray photon power law $s=-1.1$. The
locations of Beppo-SAX GRBs are shown.
GRB970228 and GRB970508 are marked with asterisks.}
\end{figure}

\begin{figure}
\epsfxsize=\hsize
\epsfbox{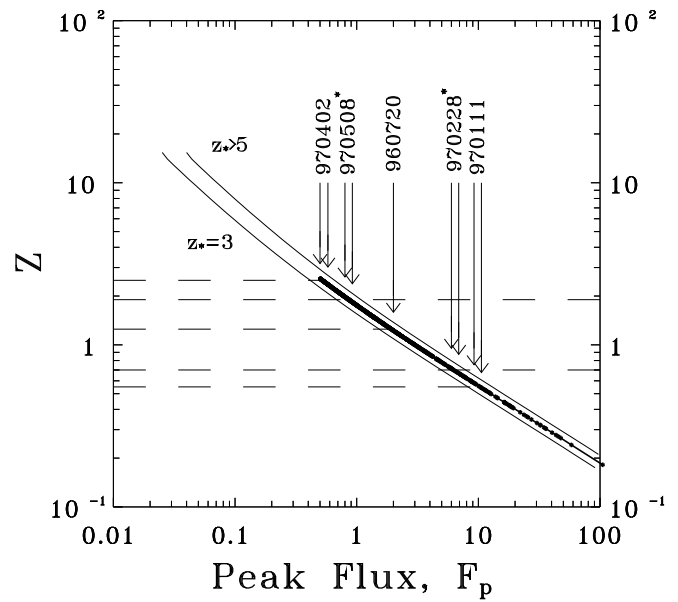}
\caption{The redshift -- peak flux dependence
in the cosmologocal models assumed for different
$z_*$ and $s=-1.1$. 3B BATSE catalog data are also plotted.
}
\end{figure}

\end{document}